\documentclass[showpacs,amsmath,amssymb,prb,twocolumn]{revtex4-1}
\usepackage{graphicx}
\usepackage{dcolumn}
\usepackage{bm}

\begin{document}

\title{Electron-vibration effects on the thermoelectric efficiency of molecular junctions}

\author{C. A. Perroni, D. Ninno, and V. Cataudella}

\affiliation{CNR-SPIN and Dipartimento di Fisica, Universita' degli Studi di Napoli ''Federico II'',\\
Complesso Universitario Monte S. Angelo, Via Cintia, I-80126 Napoli, Italy}

\begin {abstract}
The thermoelectric properties of a molecular junction model, appropriate for large molecules such as fullerenes, are studied within a non-equilibrium adiabatic approach in the linear regime at room temperature. A self-consistent calculation is implemented for electron and phonon thermal conductance showing that both increase with the inclusion of the electron-vibration coupling. Moreover, we show that the deviations from the Wiedemann-Franz law are progressively reduced upon increasing the interaction between electronic and vibrational degrees of freedom. Consequently, the junction thermoelectric efficiency is substantially reduced by the electron-vibration coupling. Even so, for realistic parameters values, the thermoelectric figure of merit can still have peaks of the order of unity. Finally, in the off-resonant electronic regime, our results are compared with those of an approach which is exact for low molecular electron densities. We give evidence that in this case additional quantum effects, not included in the first part of this work, do not affect significantly the junction thermoelectric properties in any temperature regime.
\end {abstract}

\maketitle

%\newpage

%\section{Introduction}
%\showpacs{73.23.-b, 72.15.Qm}

\section{Introduction}
In solid state systems, a voltage induces a temperature gradient and vice versa. These phenomena are known as thermoelectric effects. \cite{nolas} In some semiconducting materials, \cite{nolas,ioffe} these thermoelectric effects can be strong enough to allow either the fabrication of devices converting wasted heat into electrical energy or the realization of solid-state coolers. A key requirement to improve the energy conversion efficiency is to increase the electrical conductance ($G$) and the Seebeck coefficient ($S$) reducing the electronic ($G_K^{el}$) and lattice ($G_K^{ph}$)  contributions to the thermal conductance $G_K$. A value of the dimensionless figure of merit $ZT=G S^2 T /G_K$ of the order of $1$ is considered a fundamental prerequisite of a useful thermoelectric devices. \cite{nolas,shakouri} A clear limitation of the thermoelectric technology is that three mutually contraindicated properties of the same material have to be optimized. In  metals, for instance, $ZT$ is typically limited by the Wiedemann-Franz law, stating that the ratio  $G_K^{el}/(GT)$ is a constant (the Lorenz number) independent of the metal specificities.

In order to optimize the thermoelectric efficiency, the possibility of controlling materials at the nanoscale has been advanced. \cite{shakouri,dressel1,dressel2,koumoto} For example, large values of $ZT$  can be obtained in semiconducting nanowires with highly peaked densities of states. \cite{dressel3} It has been predicted that also molecular devices can be efficient for conversion of heat into electric energy. \cite{arad,dubi}  The improvement of thermoelectric efficiency derives from the discreteness of energy levels that leads to the violation of the Wiedemann-Franz law. \cite{Mahan}
Therefore, the emerging field of molecular thermoelectrics  can be very interesting for both the basic physics and
applications and it is mainly for this reason that the subject has attracted a lot of attention in recent years.
\cite{majum1,baheti,majum2,finch,murphy,galperin,koch,leijnse}

Single-molecule measurements have focused on the Seebeck coefficient whose sign directly provides the sign of charge carriers involved into the transport mechanisms. \cite{majum1,majum2,datta,cuevas,datta1}
Most measurements have taken into account simple small molecules where the transport is dominated by the HOMO (highest occupied molecular orbital). \cite{majum1,baheti} Recently, more controllable alignment between Fermi level and molecular orbitals (whose energy separation is still of the order of $0.5$ eV) has been achieved with larger molecules whose transport is dominated by the LUMO (lowest unoccupied molecular orbital). Measurements of conductance and Seebeck coefficient in junctions based on  $C_{60}$ molecules have been performed considering three different metallic electrodes (platinum, gold, and silver).\cite{majum2}
A very high magnitude of single-molecule thermopower has been reported ($S$ of the order of $-30$ $\mu V$/K). However, the application of a gate voltage in these kinds of measurements remains elusive. Moreover, heat transport in molecular devices remain poorly characterized owing to experimental challenges. \cite{dubi,wang1,wang2} Recently, heat transport and dissipation have been investigated in junctions with benzene-like molecules, \cite{cuevas1} even if the study has been limited to the elastic transport regime.

Non-interacting models of molecular junctions, using a Landauer-type approach, \cite{datta1} are generally not accurate since intramolecular interactions typically constitute the largest energy scales of the problem. The electron-vibration coupling, indeed, significantly affects the transport characteristics of molecular devices. \cite{cuevas,galperin1} In particular, either the molecule center of mass oscillations \cite{Park} or thermally induced acoustic phonons \cite{Qin} can be the source of coupling between electronic and vibrational degrees of freedom. The electron-vibration coupling has been studied in a fully out-of-equilibrium linear response regime with different theoretical tools, ranging from rate equation \cite{siddi,mitra,dong} (in the regime of weak tunnel coupling between molecule and electrodes) to non equilibrium Green function formalism \cite{fang,galperin2,wrudzinski} (for perturbative and intermediate to strong strength of interaction). In particular, in devices with large molecules or carbon nanotube quantum dots, the low energy of the relevant vibrational degrees of freedom has been exploited to devise a non-equilibrium adiabatic approach. This method is semiclassical for the vibrational dynamics but it is valid for arbitrary strength of electron-vibration coupling. \cite{mozy,pistol,hussein,alberto,alberto1}

The thermopower $S$ has been analyzed in a single level molecule with vibrational coupling by means of the master equation approach finding that it is sensitive to the interplay between electrons and molecular vibrations. \cite{koch} The effect of electron-vibration coupling on the thermoelectric properties of few  level molecules or dots has been studied by the rate equation \cite{leijnse} and Green function \cite{galperin,yang,zianni,ren,tagani} formalism. However, in these papers, the phonon thermal contribution $G_K^{ph}$ to the figure of merit $ZT$ has not been calculated or discussed. Recently, this contribution to the thermoelectric efficiency has been investigated in a molecular junction where a benzene molecule is directly connected to platinum electrodes. \cite{hsu} However, the electron-vibration inelastic effects on the thermoelectric properties have been evaluated only at a perturbative level.

In this paper, we have studied the thermoelectric properties of a molecular junction with electron-vibration coupling within the linear response regime focusing on the phonon thermal contribution $G_K^{ph}$ to the figure of merit $ZT$  at room temperature. The non-equilibrium adiabatic approach has been used to solve the junction model which takes into account the interplay between the low frequency center of mass oscillation of the molecule and the electronic degrees of freedom. Parameters appropriate for junctions based on $C_{60}$ molecules connected between different metallic leads have been considered. We have found that  the semiclassical $G_K^{ph}$ typically overcomes the electronic thermal conductance $G_K^{el}$, and it gets enhanced with increasing the electron-vibration coupling. Moreover, the increase of the electron-vibration coupling makes the ratio $G_K^{el}/GT$ closer to the Lorentz number. Actually, the figure of merit $ZT$ can be substantially reduced by these effects, even if, for realistic parameters of the model, it can still have peaks of the order of unity.  Finally, in the off-resonant regime, where the thermoelectric properties show peak values, we have compared the results of the adiabatic approach with those of a fully quantum formalism which is exact for low electron level density. We have stressed that, in the regime of low temperatures, the additional quantum effects in $G_K^{ph}$ not included in the adiabatic approach poorly affect the thermoelectric efficiency.

The paper is organized as follows. In Sec. II, the model of molecular junction is proposed. In Sec. III, the adiabatic approach is quickly explained.
In Sec. IV, the results within the adiabatic approach are discussed. In Sec. V, the comparison of the results within the adiabatic approach with those of a method exact in the regime of low level occupation is performed. Two Appendices close the paper: Appendix A, where the derivation of the Langevin equation for the center of mass oscillator is reported, and Appendix B, where some results about the oscillator damping rate induced by the electron-vibration coupling and position distribution function are commented.

\section{Molecular junction model}

\begin{figure}[h]
\centering
\includegraphics[scale=0.6]{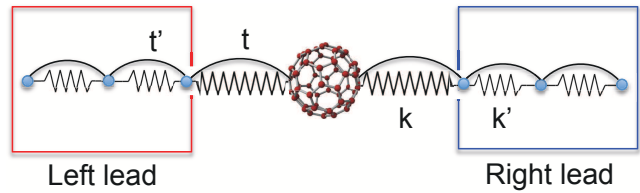}
\caption{(Color online) Sketch of the molecular junction studied in this work. The curved lines between dots (lead atoms) depict charge electron hoppings in the lead bulks ($t'$) and between lead and molecule ($t$). The broken lines between dots (lead atoms) depict springs in the lead bulks (with elastic constant $k'$) and between lead and molecule (with elastic constant $k$), therefore, they denote the phonon part. The Left Lead and the Right Lead are kept at chemical potential $\mu_L$, temperature $T_L$ and chemical potential $\mu_R$, temperature $T_R$, respectively.}
\label{pisto}
\end{figure}

The molecular junction model includes both electric and elastic coupling between leads and molecule, whereas the molecule is described as a single electron level coupled to the molecule center of mass vibration (see Fig. \ref{pisto} for a scheme of the device). The Hamiltonian $\hat{H}$ of the junction is, then, given by
\begin{equation}
\hat{H}={\hat H}_{el}+ {\hat H}_{ph} + {\hat H}_{int},
\label{Htot}
\end{equation}
where the Hamiltonian ${\hat H}_{el}$ (${\hat H}_{phon}$)  describes electronic (vibrational) degrees of freedom of both leads and molecule and ${\hat H}_{int}$ describes the coupling between electronic and vibrational degrees of freedom of the molecule. We assume that the electronic and vibrational degrees of freedom in metallic leads are not interacting \cite{cuevas,Haug}, and, therefore, the electron-vibration coupling is assumed effective only on the molecule.

The molecule is modeled as a single electronic level locally interacting with a single vibrational mode (see Fig. \ref{pisto}). This means that the focus is on a molecular electronic orbital which is sufficiently separated in energy from other orbitals. For example, in a free $C_{60}$ molecule, the LUMO energy differs from the HOMO energy for more than 1 eV. Even when  the degeneracy of the LUMO is removed by the  contact with Ag, the splitting gives rise to levels which are separated by an energy of the order of $0.5$ eV. \cite{Lu} In this last situation, the low energy level of $C_{60}$ gets aligned with the Fermi level of Ag \cite{majum2,Lu} increasing the thermopower. Finally, the choice of a single level model is motivated by the fact that a system with these electronic features can improve the thermoelectric performances. \cite{Mahan}

Since sizable figures of merit ZT are obtained for temperatures much higher than Kondo temperature of molecular junctions, \cite{cuevas} and large Seebeck coefficients are found when the molecular orbital is poorly populated, we focus our analysis in a regime where the effects of Coulomb local interactions are negligible. Hence, the electronic Hamiltonian ${\hat H}_{el}$ of Eq. (\ref{Htot}) is assumed spinless:
\begin{equation}
\hat{H}_{el}=\epsilon {\hat d^{\dag}}{\hat d}+
\sum_{q,\alpha}\varepsilon_{q,\alpha}{\hat c^{\dag}_{q,\alpha}}{\hat c_{q,\alpha}}
+ \sum_{q,\alpha} \left[ V_{q,\alpha}{\hat c^{\dag}_{q,\alpha}}{\hat d}+ h.c. \right],
\end{equation}
where the molecular electronic level has energy $\epsilon$ and  ${\hat d^{\dag}} ({\hat d})$ are creation (annihilation) operators on the molecule. The presence of a gate in the junction can be simply simulated by changing the value of the local energy $\epsilon$. \cite{cuevas}  The operators ${\hat c^{\dag}_{q,\alpha}} ({\hat c}_{q,\alpha})$ create (annihilate) electrons with momentum $q$ and energy
$\varepsilon_{q,\alpha}=\xi_{q,\alpha}-\mu_{\alpha}$ in the left ($\alpha=L$) or right ($\alpha=R$) free metallic leads.
The difference of the electronic chemical potentials in the leads, $\mu_{L}$ and $\mu_{R}$ respectively, provides
the bias voltage $V_{bias}$ applied to the junction: $\mu_{L}-\mu_{R}=e V_{bias}$, with $e$ electron charge. The left and right leads will be considered as thermostats in equilibrium at the temperatures $T_L$ and $T_R$, respectively, with temperature difference $\Delta T=T_L-T_R$ (see Fig. \ref{pisto}). Therefore, the left and right electron leads are characterized by the free Fermi distribution functions $f_{L}(\omega)$ and $f_{R}(\omega)$, respectively. The electronic tunneling between the molecular dot and a state $q$ in the lead $\alpha$
has the amplitude $V_{q,\alpha}$.  As usual for metallic leads, the density of states $\rho_{q,\alpha}$ is assumed flat about the small energy range relevant for the molecular orbital, making valid the wide-band limit: $ \rho_{q,\alpha} \mapsto \rho_{\alpha}$, $V_{q,\alpha} \mapsto V_{\alpha} $. Therefore, the full hybridization width of the molecular orbital is
$\hbar \Gamma=\sum_{\alpha } \hbar \Gamma_{\alpha}$, with $\hbar$ Planck constant and the tunneling rate $\Gamma_{\alpha}=2\pi\rho_{\alpha}|V_{\alpha}|^{2}/\hbar$. In junctions with $C_{60}$ molecules, $\hbar \Gamma$  has been estimated to be of the order of $20$ meV. \cite{natelson,mra} In the following, we consider the symmetric configuration: $\Gamma_L=\Gamma_R=\Gamma/2$; $\mu_{L}=\mu+e V_{bias}/2$, $\mu_{R}=\mu-e V_{bias}/2$, with $\mu$ average chemical potential; $T_L=T+\Delta T/2$, $T_R=T-\Delta T/2$, with $T$ average temperature.

In analogy with the electronic model, we consider only one relevant vibrational mode for the molecule. We will focus on the center of mass mode, which is expected to have the lowest frequency for large molecules. Moreover, this mode should have a frequency smaller than the Debye frequency of the metallic leads impeding the formation of localized vibrational states on the molecule. Therefore, this mode can favor the exchange of heat between the leads. Besides, in $C_{60}$ molecules, experimental results provide compelling evidence for a coupling between electron dynamics and the center of mass motion. \cite{Park}

In Eq.(\ref{Htot}), the Hamiltonian ${\hat H}_{phon}$ describes the vibrations of the molecule center of mass mode, the free phonon modes of the leads, and the coupling between them:
\begin{equation}
{\hat H}_{ph}={\hat H}_{cm} + \sum_{q,\alpha} \hbar \omega_{q,\alpha}{\hat a^{\dag}_{q,\alpha}}{\hat a_{q,\alpha}}
+ \sum_{q,\alpha} \left( C_{q,\alpha}{\hat a_{q,\alpha}}+h.c. \right) {\hat x}.\label{Hphon}
\end{equation}
The center of mass hamiltonian ${\hat H}_{cm}$ is
\begin{equation}
{\hat H}_{cm}={{\hat p}^{2}\over 2m} + \frac{k {\hat x}^{2}}{2},\label{Hcm}
\end{equation}
where ${\hat p}$ and ${\hat x}$ are the center of mass momentum and position operator, respectively, $m$ is the total large mass, $k$ is the effective spring constant, and the low frequency is $\omega_{0}=\sqrt{k/m}$. For $C_{60}$ junctions, $\hbar \omega_0$ has been estimated to be of the order of  $5$ meV. \cite{Park} Therefore, for a large molecule such as $C_{60}$, the adiabatic regime, $\omega_0 << \Gamma$, is valid.

In Eq.(\ref{Hphon}), the operators ${\hat a^{\dag}_{q,\alpha}} ({\hat a}_{q,\alpha})$
create (annihilate) phonons with momentum $q$ and frequency
$\omega_{q,\alpha}$ in the lead $\alpha$. As shown in Fig. \ref{pisto}, the left and right phonon leads will be considered as thermostats in equilibrium at the temperatures $T_L$ and $T_R$, respectively, which we assume to be the same as that of the electron leads.
In order to investigate heat exchange with the molecule, the leads phonon spectrum is assumed to be acoustic.  For silver (atomic number Z=47), gold (Z=79), and platinum (Z=78) leads considered in experimental measurements, \cite{majum2} the Debye frequency is such that $ \hbar \omega_D$ is of the order of $18.5$ meV, $15.1$ meV, and $20.7$ meV, respectively. \cite{kittel} Therefore, $ \hbar \omega_D \simeq 15-20$ meV for these leads. In any case, as for any large molecule, the center of mass mode is such that $\omega_0 << \omega_D$.

In Eq.(\ref{Hphon}), the coupling between the center of mass position and a phonon $q$ in the lead $\alpha$ is given by the elastic constant $C_{q,\alpha}$. In order to characterize this interaction, one introduces the spectral density $J(\omega)$:
\begin{equation}
J(\omega)=\frac{\pi}{2} \sum_{q,\alpha} \frac{ C_{q,\alpha}^2 } { M  \omega_{q,\alpha} }  \delta(\omega-\omega_{q,\alpha})= m \omega \tilde{\gamma}(\omega),
\label{spectral}
\end{equation}
with $M$ atomic mass of the leads and $\tilde{\gamma}(\omega)$ complex frequency dependent memory-friction kernel of the oscillator.\cite{weiss} In the regime  $\omega_0 << \omega_D$, $\tilde{\gamma}(\omega)$ can be approximated  as real and independent of frequency, providing  the damping rate $\gamma$: $\tilde{\gamma}(\omega)  \simeq \gamma$. \cite{weiss} In analogy with the electronic model, we consider the symmetric configuration: $\gamma_L=\gamma_R=\gamma/2$. If the center of mass is elastically coupled with a neighbor atom of the leads by a spring with constant $k$, one gets $\gamma \simeq 16 k^2 / (m M \omega_D^3)$. Taking the mass $m$ of the $C_{60}$ molecule and the atomic mass $M$ of Ag, Au, and Pt, $\hbar \gamma$ is of the order of $7.68$ meV, $7.74$ meV, and $2.98$ meV, respectively. The smallest value of coupling to phonon baths is due to the largest Debye frequency $\omega_D$ of platinum. In any case, $\hbar \gamma \simeq 3-8$ meV for these metals, therefore $\omega_0$ is of the same order of $\gamma$.

Finally, in Eq.(\ref{Htot}), the interaction term ${\hat H}_{int}$ derives from electrostatic interactions since it is induced by the charges injected by the leads onto the molecule. When the center of mass displacement is not large, it has been shown that this interaction is provided by a linear coupling between the electron occupation on the molecule, ${\hat n}={\hat d^{\dag}}{\hat d}$, and the $\hat{x}$ operator of the oscillator:
\begin{equation}
{\hat H}_{int}=\lambda {\hat x} {\hat n},\label{Hint}
\end{equation}
where $\lambda$ is the electron-oscillator coupling strength. \cite{alberto1} This coupling gives rise to a model known in the literature as Anderson-Holstein model. \cite{cuevas} In the following, the electron-vibration interaction will be often described in terms of the coupling energy $E_{P}=\lambda^2/(2 k)$. As reported in experimental measurements, \cite{Park} the effects of $E_P$ are not negligible in junctions with $C_{60}$ molecules.

\section{Adiabatic approach}
In this paper, we focus on the dynamics of the molecule center of mass which is generally slow for large molecules ($\hbar \omega_0 \simeq 5$ meV for junctions with $C_{60}$ molecules). Moreover, the interest is for the thermoelectric properties close to ambient temperature $T_A$. Therefore, in the following, we will study the system under arbitrary electron-vibration coupling in the range $\hbar \omega_0 \ll k_B T$ ($k_B$ Boltzmann constant) assuming that the dynamics of the center of mass can be treated as classical.

\subsection{Electron dynamics dependent on oscillator parameters}

Since the oscillator is assumed to be classic, the electronic dynamics is equivalent to that of a time dependent level with energy $E_{0}(t)=\epsilon+\lambda x(t)$, where $x(t)$ is the oscillator position.
Using the Keldysh formalism, \cite{Haug,alberto} one can exactly solve the Dyson and Keldysh equations for the molecular Green
functions integrating out exactly the lead electronic degrees of freedom.

Exploiting the fact that the full hybridization width, $ \hbar \Gamma$, between the molecule and the leads, is much larger than the oscillator energy $\hbar \omega_0$, one can determine the adiabatic expansion of the electronic Green's functions considering the explicit dependence of the electronic quantities on the oscillator position $x$ and making expansions on the oscillator velocity $v=p/m$. \cite{alberto,alberto1,alberto2,perroni,perroni1} For instance, at the lowest order, the molecular level spectral function is position dependent and given by
\begin{equation}
A(\omega,x) = \frac{\hbar \Gamma}{(\hbar \omega - \epsilon -\lambda x )^2+(\hbar \Gamma)^2/4}.
\label{function}
\end{equation}
Adiabatic expansions can be easily evaluated for all the quantities which can be expressed as a function of the level Green functions. In Appendix A,
the expansion of the level occupation $N(x,v)$ is required in order to make a self-consistent calculation of the electron and oscillator dynamics.

\subsection{Dynamics of the center of mass oscillator}

In this subsection, we analyze the slow dynamics of the molecule center of mass. The effects of the electron degrees of freedom and phonon baths
give rise to a generalized Langevin equation for the the center of mass oscillator. We are mostly interested on the junction transport properties close to room temperature, therefore in the regime $\hbar \omega_0 \ll \hbar \omega_D \simeq \hbar \Gamma < k_B T$.

The generalized Langevin equation for the oscillator dynamics is derived in Appendix A. The resulting motion equation
\begin{equation}
m \frac{d v}{d t}= F_{det}(x,v) + \xi(x,t)
\label{langevin1}
\end{equation}
has the deterministic force $F_{det}(x,v)$ and the position dependent fluctuating force $\xi(x,t)$. The deterministic force
\begin{equation}
F_{det}(x,v)=F_{gen}(x)-A_{eff}(x) v,
\label{fortot1}
\end{equation}
can be decomposed into a generalized force $F_{gen}(x)$
\begin{equation}
F_{gen}(x)=-k x +F_{\lambda}(x),
\label{fgen}
\end{equation}
where $F_{\lambda}(x)$ is due to the electron-vibration coupling,
and a dissipative force with an effective position dependent positive definite term $A_{eff}(x)$
\begin{equation}
A_{eff}(x)=A_{\lambda}(x)+m \gamma,
\label{Aeff}
\end{equation}
with $A_{\lambda}(x)$ due to the electron-vibration interaction. The fluctuating force $\xi(x,t)$ in Eq.(\ref{langevin1}) is such that
\begin{equation}
\langle \xi(x,t) \rangle=0,\;\;\;\; \langle \xi(x,t) \xi(x,t') \rangle= D_{eff}(x) \delta(t-t'), \nonumber
\label{Langevin20}
\end{equation}
where the effective position dependent noise term $D_{eff}(x)$ is
\begin{equation}
D_{eff}(x)=D_{\lambda}(x)+ k_B (T_L+T_R) m \gamma,
\end{equation}
with $D_{\lambda}(x)$ determined by the electron-vibration coupling.  In equilibrium conditions at temperature $T=T_{\alpha}$ and chemical potential $\mu$ ($V_{bias}=0$ and $\Delta T=0$), one gets a generalized fluctuation-dissipation relation $D_{eff}(x)=2 k_B T A_{eff}(x)$, since this equation is verified for each fixed position $x$.

We have numerically solved the Langevin equation under generic non-equilibrium conditions using a generalized Runge-Kutta algorithm. \cite{alberto,honey,honey1} As a result of the numerical calculations, the oscillator distribution function $Q(x,v)$ and the reduced position distribution function $P(x)$ can be determined giving insights on the oscillator positions relevant for the dynamics.

In Appendix B, we thoroughly discuss the features of the parameter $A_{\lambda}(x)/m$ in the linear response regime. The peak value of $A_{\lambda}(x)/m$ is always smaller than the values of $\gamma$ considered in this paper even for strong electron-vibration coupling $E_P$. Actually, the quantity $A_{\lambda}(x)/m$ will not strongly affect the phonon conductance $G_K^{ph}$ for realistic $E_P$ couplings. Typically, the effects due to the electron-vibration coupling on the oscillator dynamics do not represent a large perturbation with respect to those induced by the coupling to phonon leads. Obviously, as discussed in Appendix B, the effects of the electron-vibration coupling depend on the occupation of the electronic level. In the regime of low occupation, the dynamics of the oscillator is poorly influenced by these effects even for strong $E_P$.

\subsection{Electron and oscillator quantities}

In the adiabatic regime, once solved the Langevin equation, one can derive the behavior of a electronic observable $O(x,v)$ dependent on oscillator parameters. For example, we calculate a central quantity of our work, the electronic spectral function $A(\omega)$, making the average of the function $A(\omega,x)$ in Eq.(\ref{function}):
\begin{equation}
A(\omega) =\int_{-\infty}^{+\infty} d x P(x) A(\omega,x).
\label{spec}
\end{equation}

Within this approach, the features of the electronic system can be fully characterized in general non-equilibrium conditions. We point out that, within the adiabatic regime, charge and electronic energy currents are conserved on the stationary non-equilibrium states. In this paper, we will focus on the regime of linear response around the average chemical potential $\mu$ and the temperature T ($\Delta T \rightarrow 0 $, $V_{bias} \rightarrow 0$). We will evaluate the electronic conductance $G$
\begin{equation}
G=\left( \frac{e^2}{\hbar} \right) \left( \frac{\hbar \Gamma}{4} \right) \int_{-\infty}^{+\infty} \frac{ d (\hbar \omega)}{2 \pi} A(\omega) \left[ -\frac{\partial f(\omega)}{\partial (\hbar \omega)} \right],
\label{conduct}
\end{equation}
where $f(\omega)=1/(\exp{[\beta (\hbar \omega-\mu)]}+1)$ is the free Fermi distribution corresponding to the chemical potential $\mu$ and the temperature $T$, with $\beta=1/k_B T$. Then, we will calculate the Seebeck coefficient $S=-G_S/G$, with
\begin{equation}
G_S=  \left( \frac{e}{\hbar} \right) \left( \frac{\hbar \Gamma}{4 T} \right) \int_{-\infty}^{+\infty}
\frac{ d (\hbar \omega)}{2 \pi} (\hbar \omega) A(\omega) \left[ -\frac{\partial f(\omega)}{\partial (\hbar \omega)} \right].
\label{conducts}
\end{equation}
Finally, we will determine the electron thermal conductance $G_K^{el}=G_Q+T G_S S$, with
\begin{equation}
G_Q=  \left( \frac{1}{\hbar T} \right) \left( \frac{\hbar \Gamma}{4 } \right) \int_{-\infty}^{+\infty}
\frac{ d (\hbar \omega)}{2 \pi} (\hbar \omega)^2 A(\omega) \left[ -\frac{\partial f(\omega)}{\partial (\hbar \omega)} \right].
\label{conductq}
\end{equation}

Static quantities relative to the center of mass oscillator can be derived starting from the distribution functions $Q(x,v)$ and $P(x)$. However, dynamic quantities relative to the oscillator have to be inferred directly from the numerical stochastic dynamics. As discussed in Appendix A, the phonon energy currents are conserved for generic non-equilibrium states: $J_R^{ph}=-J_L^{ph}$, with $J_{\alpha}^{ph}$ current from the $\alpha$ phonon lead. Therefore, within the adiabatic approach used in this paper, the total energy conservation is satisfied since the energy currents are separately conserved for the electron and vibrational channels.
In this paper, we will focus on the phonon thermal conductance $G_K^{ph}$ calculated within the linear response regime \cite{wang3} around temperature T as
\begin{equation}
G_K^{ph}=\lim_{\Delta T \rightarrow 0} \frac{(J_L^{ph}-J_R^{ph})}{2 \Delta T}.
\end{equation}

The total thermal conductance is $G_K=G_K^{el}+G_K^{ph}$ making feasible the evaluation of the figure of merit $ZT$.
When the coupling of the center of mass mode to the metallic leads is absent ($\gamma=0$), $G_K=G_K^{el}$, so that $ZT =ZT^{el}$, which can be used to characterize the electronic thermoelectric properties.

In this paper, we will consider parameters appropriate to junctions with $C_{60}$ molecules. Therefore, $\hbar \Gamma \simeq 20$ meV will be the energy unit. As a consequence, $\Gamma$ will be the frequency unit. We will assume $\omega_0=0.25 \Gamma$ and vary $\gamma$ from $0.15 \Gamma$ to $0.40 \Gamma$ (simulating the effects of different metallic leads). We will measure lengths in units of $\lambda / k$, times in units of $1/\Gamma$, temperatures in units of $\hbar \Gamma /k_B$ (ambient temperature $T_A \simeq 1.25$ in these units). Finally, we will assume the average chemical potential $\mu=0$.

\section{Results within the adiabatic approach}

In this paper, we will discuss linear response transport properties in different conditions trying to clarify the role of the electron-vibration coupling. In particular, we will focus on the phonon energy transmission and on the electronic level variation with respect to the leads chemical potential. These variations can be controlled, in our model, changing the molecule energy level $\epsilon$ from $\epsilon=0$, that coincide with the lead chemical potential (resonant case), to a very different value $|\epsilon| >>0$ (off-resonant case).

Since our aim is to clarify the effects due to the electron-vibration coupling, in the next subsection, we start discussing, for comparison, the electronic response functions in the absence of electron-vibration interaction ($E_P=0$) and coupling to phonon leads ($\gamma=0$). In the second subsection, we will discuss the effects of electron-vibration coupling in the absence of coupling to phonon leads, then, in the final subsection, the features with full coupling.

\subsection{Results in the absence of electron-vibration interaction and coupling to phonon leads }

\begin{figure}[htb]
\centering
\includegraphics[width=9.5cm,height=8.5cm]{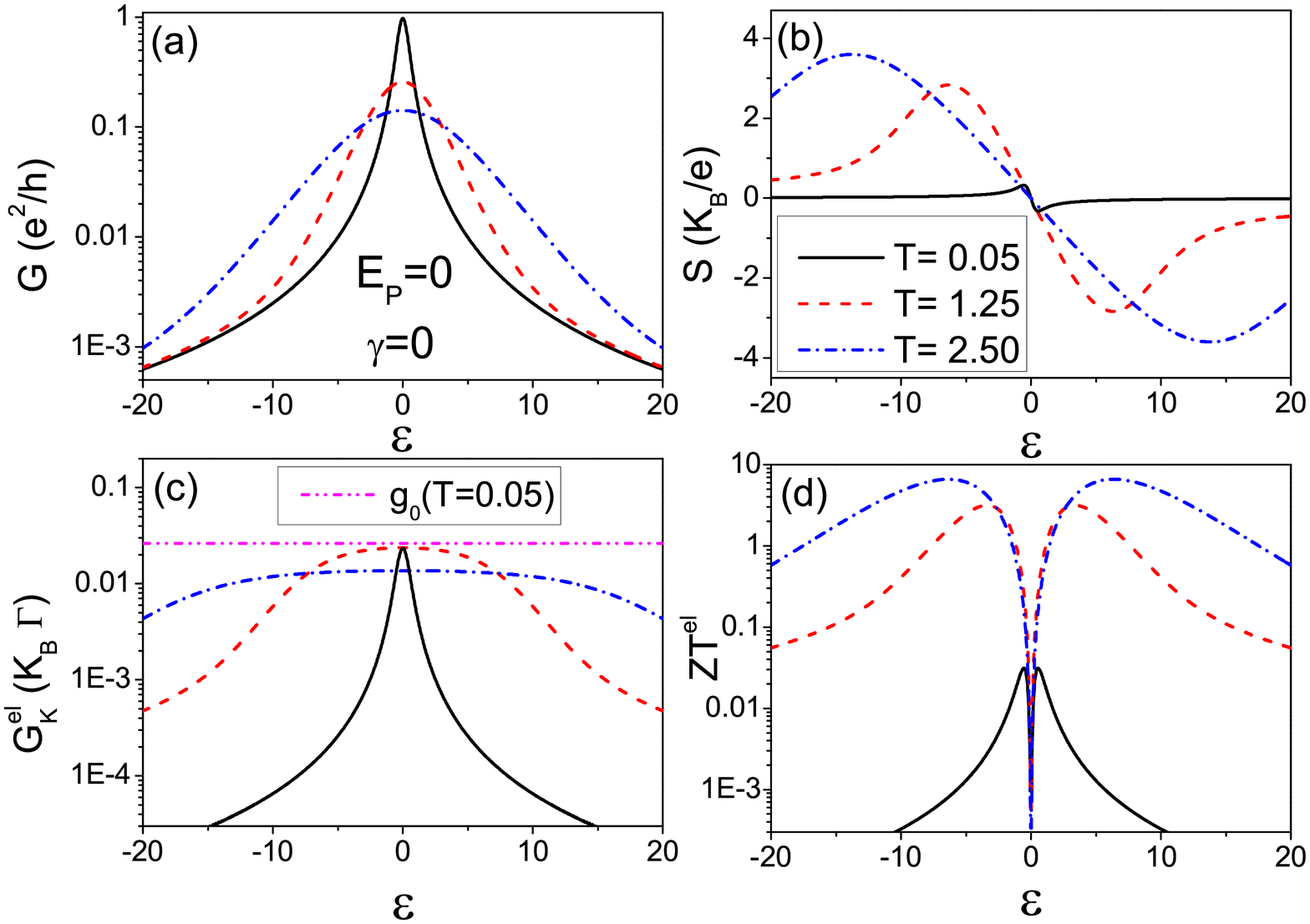}
\caption{(Color online) Panel (a): Electron conductance G (in units of $e^2/h$) as a function of the level energy (in units of $\hbar \Gamma$) for different temperatures $T$ (in units of $\hbar \Gamma /k_B$).
Panel (b): Seebeck coefficient (in units of $k_B/e$) as a function of the level energy (in units of $\hbar \Gamma$) for different temperatures $T$ (in units of $\hbar \Gamma /k_B$).
Panel (c): Electron thermal conductance $G_K^{el}$ (in units of $k_B \Gamma$) as a function of the level energy (in units of $\hbar \Gamma$) for different temperatures $T$ (in units of $\hbar \Gamma /k_B$). For comparison, the thermal conductance quantum $g_0(T) = \pi^2 k_B^2 T/(3 h)$ at $T=0.05 \hbar \Gamma /k_B$ is shown by means of the magenta double dot-dash line.
Panel (d): Electronic dimensionless figure of merit $ZT^{el}$ as a function of the level energy (in units of $\hbar \Gamma$) for different temperatures $T$ (in units of $\hbar \Gamma /k_B$).
In all the plots, electron-oscillator coupling $E_P=0$ and oscillator damping rate $\gamma=0$ (absence of coupling to phonon leads).}
\label{fig1}
\end{figure}

As shown in the panel (d) of Fig. \ref{fig1}, the thermoelectric effect is maximum at specific off-resonant conditions  where $ZT^{el}$  can reach values of the order of unity or even larger, depending on temperature. In fact, we have analyzed the behavior of transport properties for different temperatures finding that $ZT^{el}$ gets larger with increasing temperature. These behaviors result from the relevant role played by the Seebeck coefficient for these regimes of parameters.

In the off-resonant regime, the charge conductance is expected to be small. As shown in the panel (a) of Fig. \ref{fig1}, for $|\epsilon| >>0$, the conductance is much smaller than the conductance quantum $e^2/h$ ($e^2/h$ is about $3.87 \times 10^{-5}$ S). In particular, for $\epsilon=20$, $G$ is of the order of $10^{-3}$ $e^2/h$ in agreement with the order of magnitude of experimental data in $C_{60}$. \cite{majum2} On the other hand, in the resonant case $\epsilon=0$, the conductance reaches the quantum at low temperatures and, then, rapidly decreases to values of the order of $0.1$ $e^2/h$ with increasing temperature.

As shown in the panel (c) of Fig. \ref{fig1}, for both resonant and off-resonant conditions, $G_K^{el}$ is small when measured in units of $k_B \Gamma$  ($k_B \Gamma$ is about $419.8$ pW/K for $\hbar \Gamma \simeq 20$ meV). In the resonant case, $G_K^{el}$ decreases with increasing temperature and it is of the order of $0.01-0.02 k_B \Gamma \simeq 4-8$ pW/K for different temperatures, therefore it is smaller than the molecular conductance (per chain) of the order of $50$ pW/K measured in hydrocarbon molecules anchored to a gold substrate. \cite{wang2} Next, we will stress that the electron-vibration coupling will induce an increase of the thermal conductivity providing results compatible with experimental estimates. At low temperatures ($T=0.05 \hbar \Gamma /k_B$ is about $12$ K), in the resonant case, the result practically coincides with the thermal conductance quantum $g_0(T) = \pi^2 k_B^2 T/(3 h)$ at that temperature ($g_0(T) \simeq  9.456 \times 10^{-13} (W/K^2) T)$. \cite{jezouin} In the off resonant regime  $|\epsilon| >>0$, $G_K^{el}$ gets larger with increasing temperature, but, in the investigated temperature range, it is always smaller than $0.01 k_B \Gamma$.

As shown in the panel (b) of Fig. \ref{fig1}, the Seebeck coefficient $S$ shows large variations that are responsible for the behavior of $ZT^{el}$ shown in panel (d) of Fig.(\ref{fig1}). At low temperatures and in quasi-resonant regime ($\epsilon \simeq 0$), $S$ is very small whereas, in the off-resonant condition and at high temperatures, the thermopower can be very large. When $\epsilon$ is positive (n-type behavior), $S$ is negative. In particular, for $\epsilon=20$, $S$ is about $-0.45 k_B/e \simeq - 38.5 \mu V/K$ in agreement with the magnitude of experimental data in $C_{60}$. \cite{majum2} On the other hand, when $\epsilon$ is negative (p-type behavior), $S$ is positive. Therefore, as expected, the sign of $S$ is sensitive to the charges responsible for the transport. In the off-resonant case, the peak values of $S$ are quite large since they are of the order of a few $k_B/e$ ($k_B/e$ is about $86$ $\mu$ eV/K).

Summarizing, in the off-resonant condition and at high temperatures, the reduction of the conductance $G$ is fully compensated by the strong increase of the Seebeck coefficient $S$. Moreover, for $\gamma=0$, the thermal conductance is small. Therefore, as shown in the panel (d) of Fig. \ref{fig1}, $ZT^{el}$ can acquire values larger than $1$. Finally, we stress that the peak values of $ZT^{el}$ at room temperature are almost coincident with maxima and minima of the Seebeck coefficient $S$.

\subsection{Results in the absence of coupling to phonon leads }

\begin{figure}[h]
\centering
\includegraphics[width=8.5cm,height=8.5cm]{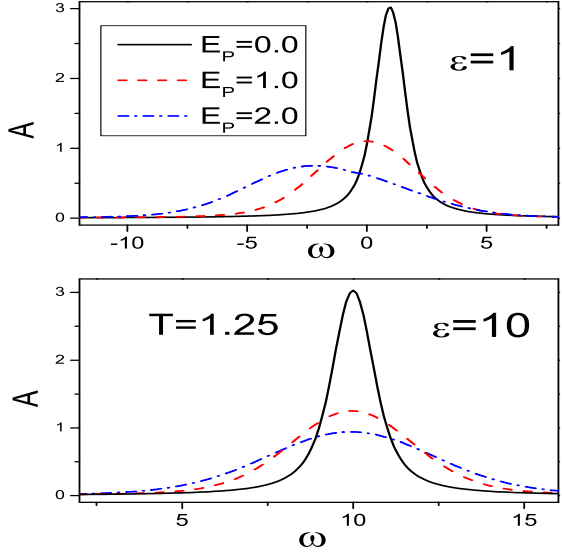}
\caption{(Color online) Spectral function (in units of $1/\hbar \Gamma$) as a function of the frequency $\omega$ (in units of $\Gamma$) for different values of $E_P$ (in units of $\hbar \Gamma$) at $\epsilon=1 \hbar \Gamma$ (Upper Panel) and $\epsilon=10 \hbar \Gamma$ (Lower Panel).
In all the plots, $T=1.25 \hbar \Gamma \ k_B$ (close to room temperature), $\omega_0=0.25 \Gamma$, and oscillator damping rate $\gamma=0$ (absence of coupling to phonon leads).}
\label{fig2a}
\end{figure}

In this subsection, we focus on the effects of the electron-vibration coupling on the electronic response functions still in the absence of coupling to phonon leads ($\gamma=0$). We focus on a temperature close to room temperature ($T=1.25$).

First, we analyze the behavior of the spectral function as a function of the electron-vibration coupling. In the upper panel of Fig. \ref{fig2a},  we show the spectral function for different values of the electron-vibration coupling in the quasi-resonant case $\epsilon=1$. We point out that there is a strong transfer of spectral weight toward low frequency with increasing $E_P$. In particular, at $E_P=1$, the spectral function is peaked on the chemical potential $\mu=0$. Indeed, when $\epsilon=E_P$, one gets the level density $n=0.5$, that is, the half-filling condition occurs. Moreover, at $E_P=2$, the peak is much smaller in frequency, and the spectral function acquires asymmetric features. As shown in the lower panel of Fig. \ref{fig2a}, a different behavior takes place in the off-resonant regime ($\epsilon=10$ is considered in the figure). For the considered values of $E_P$, the spectral function gets enlarged, but its peak position is quite rigid. Therefore, the behavior of the spectral function is different in the resonant and off-resonant regime for fixed values of $E_P$.

\begin{figure}[h]
\centering
\includegraphics[width=9.5cm,height=8cm]{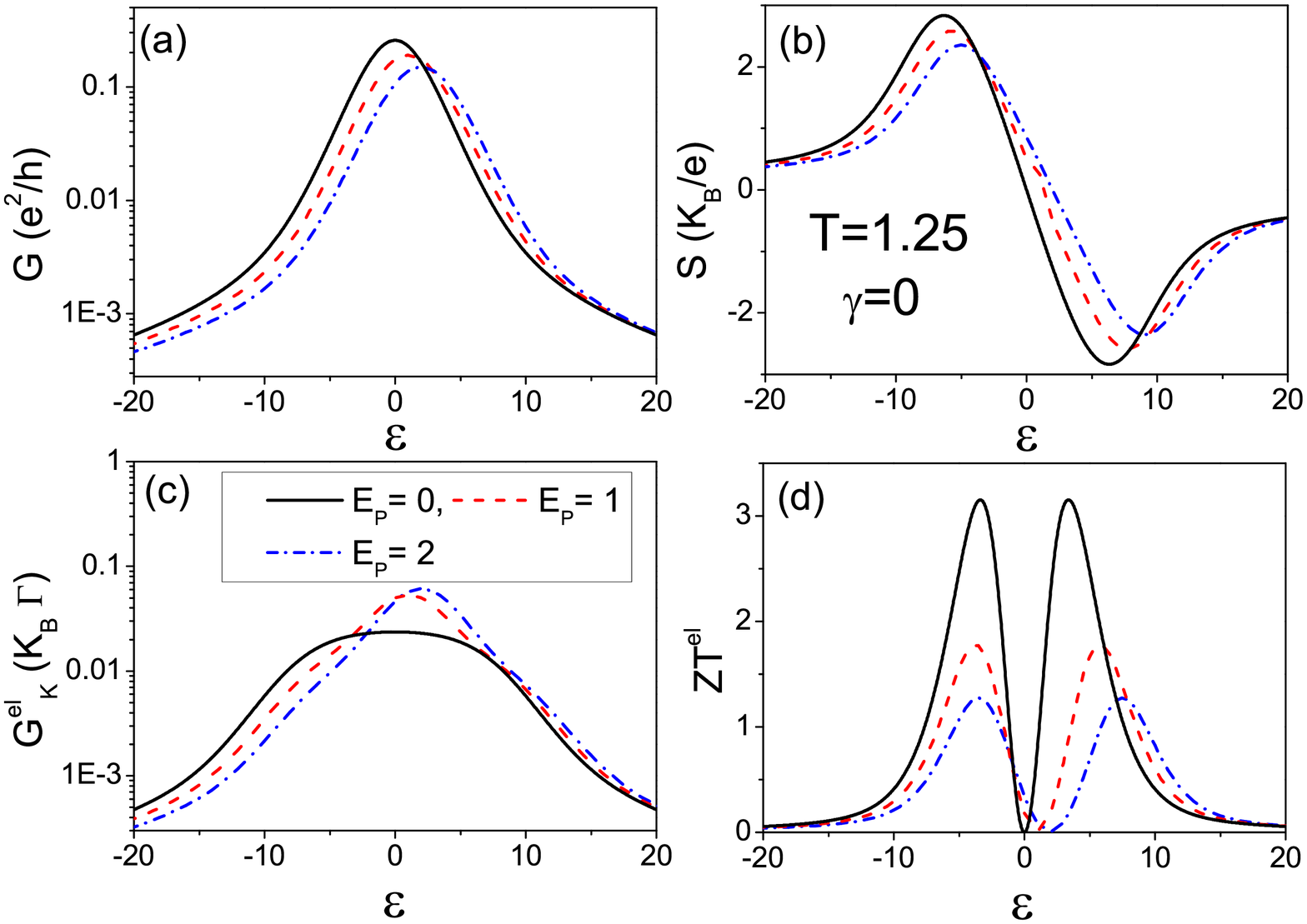}
\caption{(Color online) Panel (a): Electron conductance G (in units of $e^2/h$) as a function of the level energy (in units of $\hbar \Gamma$) for different values of $E_P$ (in units of $\hbar \Gamma$).
Panel (b): Seebeck coefficient (in units of $k_B/e$) as a function of the level energy (in units of $\hbar \Gamma$) for different values of $E_P$ (in units of $\hbar \Gamma$).
Panel (c): Electron thermal conductance $G_K^{el}$ (in units of $k_B \Gamma$) as a function of the level energy (in units of $\hbar \Gamma$) for different values of $E_P$ (in units of $\hbar \Gamma$).
Panel (d): Electronic dimensionless figure of merit $ZT^{el}$ as a function of the level energy (in units of $\hbar \Gamma$) for different values of $E_P$ (in units of $\hbar \Gamma$).
In all the plots, $T=1.25 \hbar \Gamma \ k_B$ (close to room temperature), and oscillator damping rate $\gamma=0$ (absence of coupling to phonon leads).}
\label{fig2}
\end{figure}

The features of the spectral function affect the behavior of the thermoeletric properties. As shown in Fig. \ref{fig2}, the most relevant effect of the coupling $E_P$ on the conductance $G$ (panel (a)) and the Seebeck coefficient $S$ (panel (b)) is to shift the curves and reduce the magnitude of the response function. The shift of the conductance peak and of the value where the Seebeck coefficient vanishes is given by $E_P$ ($n=0.5$ for $\epsilon=E_P$). We point out that, at fixed level energy, unlike the conductance $G$, the Seebeck coefficient shows a large sensitivity to the change of the coupling $E_P$. For example, this occurs for energies close to the minimum and the maximum. For larger values of $\epsilon$, there is an inversion in the behavior of $S$ with increasing the coupling $E_P$.

A different behavior is shown by the thermal conductance $G_K^{el}$ (panel (c) of Fig. \ref{fig2}). Indeed, $G_K^{el}$ can be enhanced with increasing the electron-oscillator coupling $E_P$. For example, at the resonance and $E_P=1$, $G_K^{el}$ is of the order of $0.05 k_B \Gamma \simeq 20$ pW/K for different temperatures, therefore it gets closer to the conductance of the order of $50$ pW/K measured in hydrocarbon molecules. \cite{wang2} Next, we will see that the total conductance, including also the phonon contribution, is even closer to these experimental data.
Actually, within the adiabatic approach, more energetic channels open with increasing the electron-vibration coupling since the effective level is renormalized by the oscillator dynamics that becomes less localized.  The reduction of $G$ and $S$ combined with the enhancement of $G_K^{el}$ leads to a sensible reduction of the figure of merit $ZT^{el}$ with increasing $E_P$. For $E_P=0$, the peak value of $ZT^{el}$ is around $3$, while, for $E_P=1$, the peak value is smaller than 2. Therefore, even if one neglects the role of phonon thermal conductance, the electron-vibration coupling is able to induce an important reduction of the figure of merit.

\subsection{Results in the case of full coupling}

\begin{figure}[h]
\centering
\includegraphics[width=7.5cm,height=9cm]{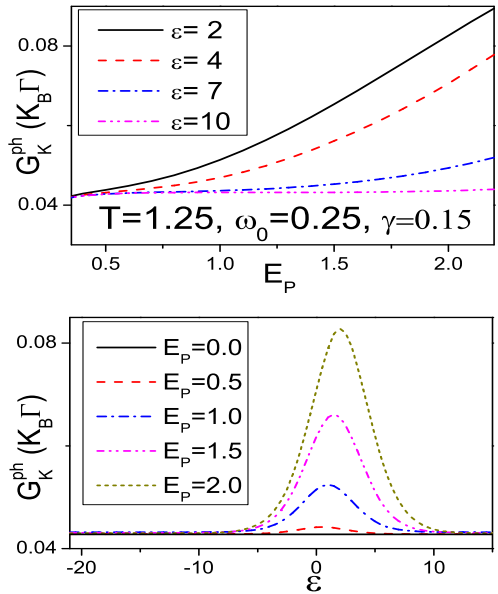}
\caption{(Color online) Upper Panel:  Phonon thermal conductance $G_K^{ph}$ (in units of $k_B \Gamma$) as a function of electron-vibration coupling $E_P$ (in units of $\hbar \Gamma$) for different values of level energy $\epsilon$ (in units of $\hbar \Gamma$).
Lower Panel: Phonon thermal conductance $G_K^{ph}$ (in units of $k_B \Gamma$) as a function of level energy $\epsilon$ (in units of $\hbar \Gamma$) for different values of electron-vibration coupling $E_P$ (in units of $\hbar \Gamma$).
In all the plots, $T=1.25 \hbar \Gamma / k_B$ (close to room temperature), the oscillator damping rate $\gamma=0.15 \Gamma$, and $\omega_0=0.25 \Gamma$.}
\label{fig3}
\end{figure}

Finally, we analyze the case when also the coupling of the center of mass mode to the metallic leads is present. First, as shown in Fig. \ref{fig3}, we focus on the phonon thermal conductance $G_K^{ph}$. We start from the minimum value of $\gamma$ ($\gamma \simeq 0.15 \Gamma$) considered in this work. We find that, for weak electron-vibration coupling $E_P$ (see upper panel of Fig. \ref{fig3}) or in the off-resonant regime $|\epsilon| \gg 0$ (see lower panel of Fig. \ref{fig3}), $G_K^{ph}$ reaches its lowest value that is close to $0.04 k_B \Gamma \simeq 16$ pW/K,  the numerical value  obtained when  electron-vibration effects are neglected (see the analytical estimate of $G_K^{ph}$ given in Eq.(\ref{gkfon}) of Appendix A). This value corresponds only to the contribution given by the  phonon leads. We point out that this asymptotic value of $G_K^{ph}$ is always larger than the values of $G_K^{el}$ shown in Fig. \ref{fig1} (corresponding to $E_P=0$). Therefore, $G_K^{ph}$ plays the major role in determining the total thermal conductance $G_K$ for weak electron-vibration coupling.

In the lower panel of Fig. \ref{fig3}, we show that $G_K^{ph}$ always gets larger with increasing the electron-vibration coupling $E_P$. Actually, the electron-oscillator coupling gives rise to an additional damping rate whose effect is to enhance the thermal conductivity $G_K^{ph}$ (see also Appendix A).
In the quasi-resonant regime, the increase of $G_K^{ph}$ can be also favored by a softening of the oscillator frequency \cite{alberto}.
Moreover, we notice that the conductances $G_K^{el}$ and $G_K^{ph}$ tend to acquire similar values with increasing the electron-vibration coupling. For example, at the resonance and $E_P=1.0$, $G_K^{el}$ and $G_K^{ph}$ are both of the order of $0.05 k_B \Gamma \simeq 20$ pW/K, therefore the total conductance is of the order of $0.1 k_B \Gamma \simeq 40$ pW/K, a value perfectly compatible with the conductance of the order of $50$ pW/K measured in hydrocarbon molecules. \cite{wang2} Finally, if one considers larger values of $\gamma$ (for example $\gamma \simeq 0.4 \Gamma$), $G_K^{ph}$ plays an even more important role in $G_K$.

As discussed above and in Appendix B, the effects of the electron-vibration coupling on the oscillator dynamics depends not only on the strength of the coupling $E_P$, but also on the occupation of the electronic level.  Actually, the behavior of  $G_K^{ph}$ is strongly dependent on the value of level energy $\epsilon$. As shown in the upper panel of Fig. \ref{fig3}, in the quasi-resonant case ($\epsilon=2$), the increase of $G_K^{ph}$ as a function of the electron-vibration coupling $E_P$ is marked. Actually, for $E_P=2$, the value of $G_K^{ph}$ is doubled. On the other hand, in the off-resonant regime of low level occupation, the dynamics of the oscillator is poorly influenced by the electron-vibration effects, even if $E_P$ is not small. Finally, in the lower panel of Fig. \ref{fig3}, we have analyzed the behavior of $G_K^{ph}$ as a function of the level energy $\epsilon$ for different values of $E_P$. As expected, $G_K^{ph}$ shows the largest deviations from the asymptotic value in the
quasi-resonant case. We point out that the peak value is practically coincident with the value of $E_P$, therefore, $G_K^{ph}$ is strongly sensitive to the renormalizations of the electron level induced by the electron-vibration coupling.

\begin{figure}[h]
\centering
\includegraphics[width=7cm,height=10.0cm]{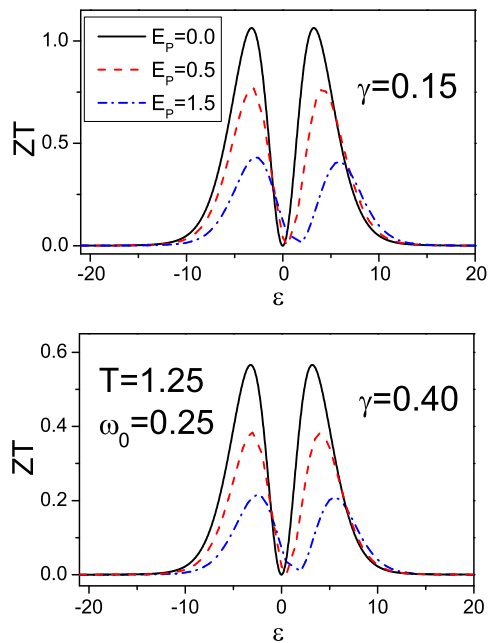}
\caption{(Color online) Dimensionless figure of merit $ZT$ as a function of level energy $\epsilon$ (in units of $\hbar \Gamma$) for different values of electron-vibration coupling $E_P$ (in units of $\hbar \Gamma$) at $\gamma=0.15 \Gamma$ (Upper Panel) and $\gamma=0.40 \Gamma$ (Lower Panel). In all the plots, $T=1.25 \hbar \Gamma / k_B$ (close to room temperature), and $\omega_0=0.25 \Gamma$. }
\label{fig4}
\end{figure}

In Fig.\ref{fig4}, we focus on the figure of merit $ZT$ for different values of electron-vibration coupling $E_P$ at $\gamma=0.15 \Gamma$ (upper panel) and $\gamma=0.40 \Gamma$ (lower panel). From the comparison with the results shown in Figs.\ref{fig1} and \ref{fig2}, we stress that the role of the phonon thermal conductance $G_K^{ph}$  is important in inducing a suppression of $ZT$. For example, already at $E_P=0$, the peak value of $ZT$ is decreased by a factor of $3$ for $\gamma=0.15 \Gamma$ and a factor of $5$ for $\gamma=0.40 \Gamma$. Obviously, the electron-vibration coupling provides an additional decrease. However, in the intermediate coupling regime ($E_P=0.5$ in Fig.\ref{fig4}), the reduction of $ZT$ due to the electron-vibration coupling is not strong. Only in the strong coupling regime, $ZT$ acquires peak values less than half unity. Finally, we stress that, in any coupling regime,  the peak values of $ZT$ are always linked to the maxima and the minima of the Seebeck coefficient. Summarizing, the cooperative effects of phonon leads and electron-vibration coupling on the molecule are able to weaken the thermoelectric performance of this kind of device. However, within a realistic weak to intermediate electron-vibration coupling regime, the figure of merit $ZT$ is still of the order of unity, making these devices useful for thermoelectric applications.

\begin{figure}[h]
\centering
\includegraphics[width=7.5cm,height=9.0cm]{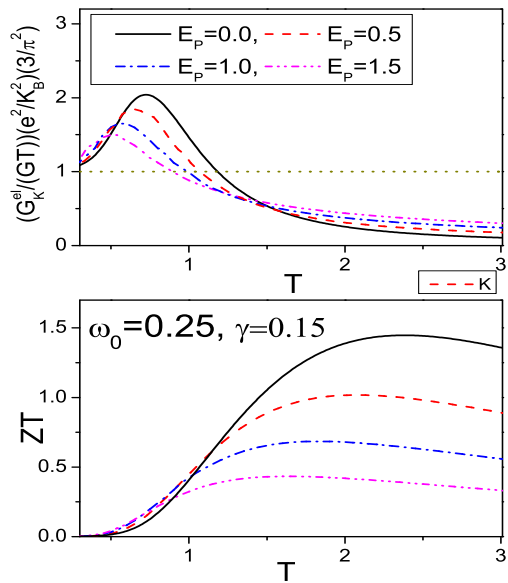}
\caption{(Color online)  Ratio $G_K^{el}/(GT)$ in units of the Lorenz number $L=\pi^2 k_B^2/(3 e^2)$ (Upper Panel) and dimensionless figure of merit $ZT$ (Lower Panel) as a function of temperature (in units of $\hbar \Gamma / k_B$) for different values of electron-vibration coupling $E_P$ (in units of $\hbar \Gamma$). In all the plots, $\gamma=0.15 \Gamma$ and $\omega_0=0.25 \Gamma$.}
\label{fig5}
\end{figure}

As shown in Fig.\ref{fig1}, the temperature plays an important role in enhancing the Seebeck coefficient $S$, and, consequently, $ZT$. Therefore, in Fig.\ref{fig5}, we study the temperature behavior of the response functions. In the upper panel of Fig.\ref{fig5}, we report the ratio $G_K^{el}/(GT)$ in units of the Lorenz number $L=\pi^2 k_B^2/(3 e^2)$ (L is about $2.44 \times 10^{-8}$ W $\Omega K^{-2}$).  When this ratio is one (dot line in the upper panel of Fig.\ref{fig5}), the Wiedemann-Franz law is satisfied. We stress that, in this system, this law is followed only at very low temperatures. \cite{murphy} Indeed, with increasing temperature, already at $E_P=0$, the Wiedemann-Franz law is violated in our system due to a peaked density of states. We point out that, as reported in the upper panel of Fig.\ref{fig5}, the violation of this law as a function of temperature is reduced with increasing the coupling $E_P$.

It has been suggested that the violation of the Wiedemann-Franz law favors the increase of the figure of merit $ZT$ in system with discrete density of states. \cite{Mahan} Actually, as shown in the lower panel of Fig.\ref{fig5}, the largest $ZT$ are present for $E_P=0$. Besides, the figure of merit $ZT$ shows a maximum around $1.5$ close to a temperature of the order of $2 T_A$. This structure is maintained with increasing the electron-vibration coupling. In particular, for the intermediate coupling $E_P=0.5$, the maximum value for $ZT$ is around $1$. Therefore, the thermoelectric performances of the junction can be still optimized by varying the temperature.

\section{Comparison with an approach exact in the regime of low level occupation}

The regime of low level occupation is the most important to get large Seebeck coefficients $S$ and, consequently, the figure of merit. In the off-resonant regime $|\epsilon| \gg 0$, the electronic Green functions of the model can be exactly calculated if, as assumed in this paper, the wide band limit is used for the leads. \cite{Mahan1,almb} This calculation takes fully into account the quantum nature of the oscillator, so that it is valid also at very low temperatures ($T \le \omega_0$). We stress that, within the off-resonant regime, the oscillator dynamics is very weakly perturbed by the effects of the electron-vibration coupling, but it is very sensitive to the coupling to phonon leads.

We focus on the retarted Green function in real time $G_{R}(t)$ at the average temperature $T$
\begin{equation}
G_{R}(t)=G_{R}^{(0)}(t) \exp{\left[ - \frac{1}{2} E_P^2   \phi(t) \right]},
\label{gret}
\end{equation}
which is non-perturbative in the electron-vibration coupling, as it is in the adiabatic approach. Indeed, in Eq.(\ref{gret}), $G_{R}^{(0)}(t)$ is the Green function without electron-vibration effects
\begin{equation}
G_{R}^{(0)}(t)=- \frac{i}{\hbar} \theta(t) \exp{ \left[ -i t \left( \frac{\epsilon}{\hbar} - i \Gamma/2 \right) \right]},
\end{equation}
with $\theta(t)$ Heaviside function, and $\phi(t)$ is the phonon term
\begin{equation}
\phi(t)=i \int_0^t d t_1 \int_0^t d t_2 D(t_1-t_2),
\label{firet}
\end{equation}
with $D(t)$ phonon Green function. Due to the coupling of the oscillator to phonon leads, $D(t)$ is not the free Green function, but contains the damping $\gamma$. At finite temperature $T$, $D(t)$  is
\begin{equation}
D(t)=\theta(t) D_1(t)+ \theta(-t) D_2(t)
\end{equation}
where $D_1(t)$ is
\begin{equation}
D_1(t)=- \frac{i}{\hbar} \left[ (N_0+1) \exp{(-i \tilde{\omega}_0 t)}+ N_0 \exp{(i \tilde{\omega}_1 t)}  \right]
\end{equation}
and $D_2(t)$
\begin{equation}
D_2(t)=- \frac{i}{\hbar}  \left[ (N_0+1) \exp{(i \tilde{\omega}_0 t)}+ N_0 \exp{(-i \tilde{\omega}_1 t)}  \right],
\end{equation}
with $N_0$ Bose distribution function at temperature $T$, $\tilde{\omega}_0=\omega_0-i \gamma/2$, and $\tilde{\omega}_1=\omega_0+i \gamma/2$. Upon taking the Fourier transform of $G_{R}(t)$ in Eq.(\ref{gret}), we calculate the spectral function. Using this spectral function, we can evaluate the electron transport properties through Eqs.(\ref{conduct}),(\ref{conducts}), and (\ref{conductq}) in analogy with the adiabatic approach.

\begin{figure}[h]
\centering
\includegraphics[width=7.5cm,height=8.0cm]{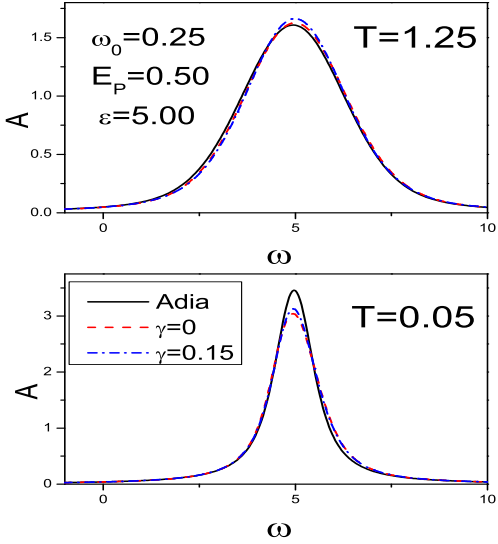}
\caption{(Color online) Spectral function as a function of the frequency (in units of $\Gamma$) for adiabatic approach (solid line) and low density approach (dashed line for $\gamma=0$, dash-dotted line for $\gamma=0.15$) at temperature $T=1.25 \hbar \Gamma / k_B$ (Upper Panel) and temperature
$T=0.05 \hbar \Gamma / k_B$ (Lower Panel). In all the plots, $E_P=0.5 \hbar \Gamma$, $\epsilon=5 \hbar \Gamma$, and $\omega_0=0.25 \Gamma$.}
\label{fig6}
\end{figure}

The spectral function calculated here will be compared with that obtained within the adiabatic approach in Eq.(\ref{spec}). First, in the upper panel of Fig. \ref{fig6}, we consider the spectral functions at $T=1.25 \hbar \Gamma / k_B$, which is close to room temperature. Moreover, we consider the off-resonant regime $\epsilon=5$, which is very close to the minimum of the Seebeck coefficient. The spectral weight up to $0$ (position of the chemical position) indicates that the level occupation $n$ is small (less than $0.1$). The agreement between the spectral functions calculated within the two approaches is excellent. The peak positions for both approaches are at $\omega=\epsilon$ and the widths of the curves perfectly match. The role of $\gamma$ is not relevant, since, in any case, it is much smaller than $\Gamma$. Obviously, with decreasing the temperature, the two approaches tend to differ. In the lower panel of Fig.\ref{fig6}, we have considered the worst case of very low temperature ($T=0.05 \hbar \Gamma / k_B$). We point out that the agreement between the two approaches is still good. Indeed, the approach exact at low molecule occupation slightly favors a small transfer of  spectral weight at high frequency. In any case, strong similarities in the spectral function will give rise to analogous behaviors of electron transport properties within the two approaches. \cite{wingreen}

In the regime of low molecule occupation, the phonon thermal conductance $G_K^{ph}$ at the average temperature $T$ can be calculated neglecting all the renormalization effects due to electron-vibration coupling but retaining all the quantum contributions \cite{wang4,pana}:
\begin{equation}
G_K^{ph}=\frac{k_B \gamma^2}{2} \int_{-\infty}^{+\infty} \frac{ d \omega}{2 \pi}
\frac{\omega^2}{(\omega_0^2-\omega^2)^2+(\omega \gamma)^2} F(\omega),
\label{condq}
\end{equation}
with $F(\omega)$ given by
\begin{equation}
F(\omega)=\left( \frac{\hbar \omega}{k_B T} \right)^2 \frac{ \exp{\left( \frac{\hbar \omega}{k_B T} \right) } }
{\left[ \exp{\left( \frac{\hbar \omega}{k_B T} \right) } -1 \right]^2}.
\end{equation}

\begin{figure}[h]
\centering
\includegraphics[width=9.5cm,height=7.5cm]{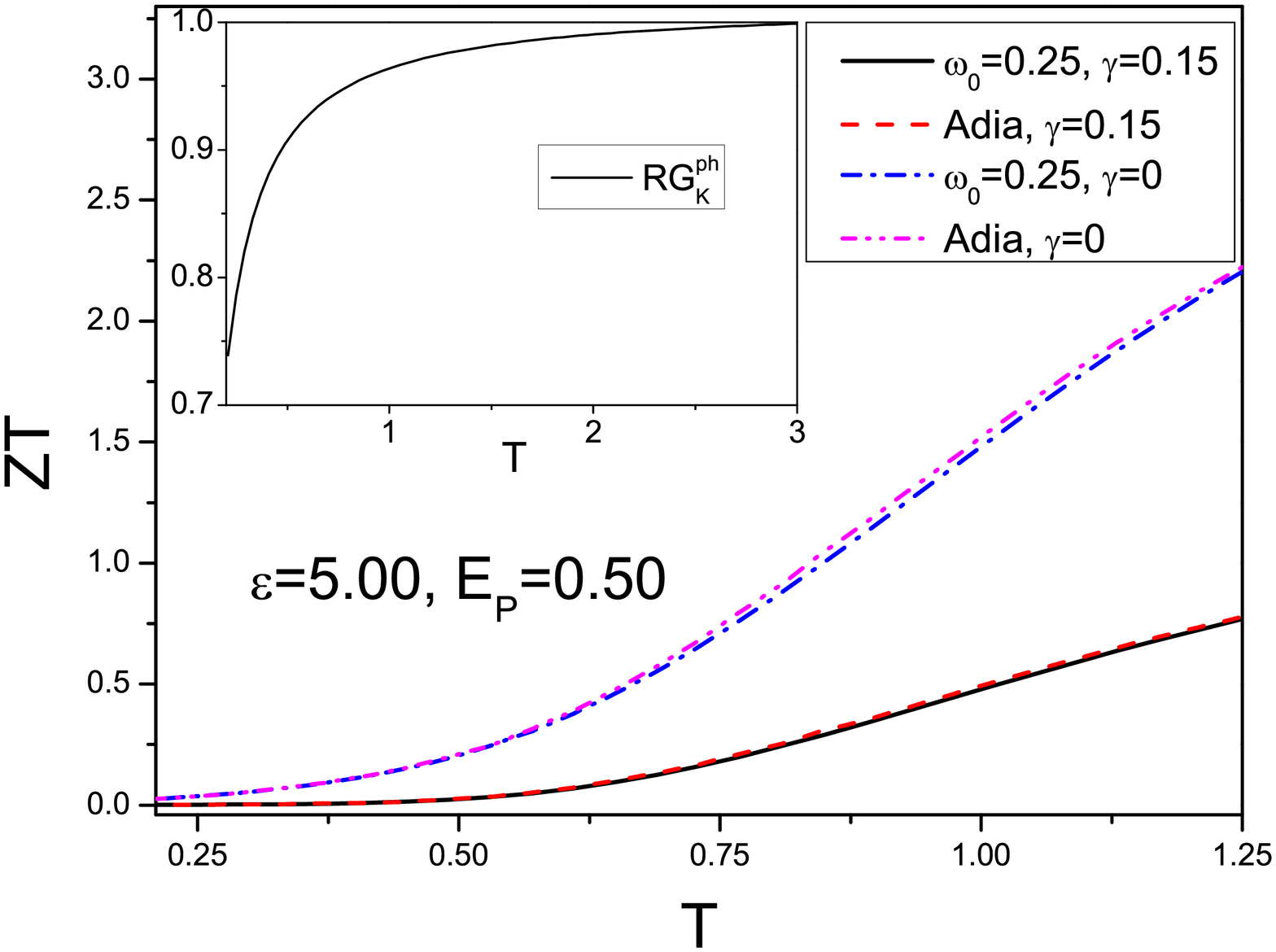}
\caption{(Color online) Figure of merit $ZT$ as a function of temperature (in units of $\hbar \Gamma / k_B$) for different values of parameters and approaches. Solid line: low density quantum approach for $\gamma=0.15 \Gamma$; Dash line : adiabatic approach for $\gamma=0.15 \Gamma$; Dash-dot line: low density quantum approach for $\gamma=0$; Dash-double dot line : adiabatic approach for $\gamma=0$. In the figure, $E_P=0.5 \hbar \Gamma$, and $\epsilon=5 \hbar \Gamma$.}
\label{fig7}
\end{figure}

The quantity $RG_K^{ph}$ will indicate the ratio between the conductance in Eq.(\ref{condq}) and that calculated in Eq. (\ref{gkfon}) of Appendix A based on the adiabatic semiclassical approach. In the off-resonant case corresponding to $\epsilon=5 \hbar \Gamma$, the effects due to electron-vibration coupling
$E_P=0.5 \hbar \Gamma$ on the oscillator dynamics are negligible. Due to the quantum effects, $F(\omega)$ is smaller than unity implying that the ratio $RG_K^{ph}$ shares the same behavior. In the inset of Fig. \ref{fig7}, we report this ratio showing that it goes rapidly to 1 with increasing temperature. Actually, at $T= \hbar \Gamma / k_B $, this ratio is already around $0.95$. Quantum corrections are relevant only for temperatures of the order of $\hbar \omega_0 / k_B$ (much smaller than
$\hbar \Gamma / k_B$).

Finally, we have compared the figure of merit calculated within the adiabatic semiclassical approach with that determined in this section which retains quantum corrections. We still focus on the off-resonant regime corresponding to $\epsilon=5$, neglecting the effects of the electron-vibration coupling on the phonon thermal conductance (we use Eq. (\ref{condq}) within the low density quantum approach). At very low temperatures, the figure of merit is definitely small. As reported in Fig. \ref{fig7}, $ZT$ rapidly gets larger with increasing temperature. First, we have compared $ZT$ for the two approaches in the low temperature range at $\gamma=0$. We find that, even in this regime, the agreement between the two approaches is good. $ZT$ calculated in the adiabatic approach is slightly larger than that obtained in the low density approach. Then, we have considered $ZT$ for the two approaches at $\gamma=0.15 \Gamma$. Therefore, we include the contribution from the phonon thermal conductance $G_K^{ph}$. Since, as shown in the inset of  Fig. \ref{fig7}, $G_K^{ph}$ with quantum terms in Eq.(\ref{condq}) is smaller than the adiabatic semiclassical quantity, the $ZT$ calculated within the two approaches perfectly match. Therefore, the thermoelectric properties discussed in this paper are consistently described also in the low temperature regime. The comparison between two approaches cannot be extended in the regime of high temperatures since the level density $n$ increases making the quantum low density approach less valid. For $\epsilon > 5$, the range of temperature for the comparison increases finding a perfect agreement between the two approaches.

\section{Conclusions}

In this paper, the thermoelectric properties of a molecular junction have been studied within the linear response regime  at room temperature. In particular, we have analyzed the role played by the phonon thermal contribution $G_K^{ph}$  on the figure of merit $ZT$ in the presence of electron-vibration coupling. The interplay between the low frequency center of mass oscillation of the molecule and the electronic degrees of freedom has been investigated using a non-equilibrium adiabatic approach. Parameters appropriate to $C_{60}$ molecules connected with different metallic leads have been considered. The semiclassical $G_K^{ph}$ is typically of the same order of or larger than electronic thermal conductance $G_K^{el}$. Both conductances are affected by the changes in the occupation of electron levels, and they get larger with increasing the electron-vibration coupling. Moreover, deviations from the Wiedemann-Franz law are progressively reduced with increasing the electron-vibration coupling. Therefore, the figure of merit $ZT$ depends appreciably on the behavior of $G_K^{ph}$ and electron-vibration coupling. Indeed, for realistic parameters of the model, $ZT$ can be substantially reduced, but it can still have peaks of the order of unity with enhancements due to temperature increase. Finally,  we have compared the results of the adiabatic approach with those of a formalism which is exact for low electron level density. We have pointed out that the additional quantum effects included in $G_K^{ph}$ poorly influence the thermoelectric properties in any regime of temperatures.

The nanoscopic junction investigated in this paper is advantageous compared to bulk or other low-dimensional structures in providing a mechanism to keep the phonon thermal conduction low. Actually, the enhancement of $G_K^{ph}$ due to the electron-vibration coupling at most provides a factor of two to a value that is small compared to bulk conductances. The phonon thermal conductance depends not only on the properties of the metallic leads, but also of the tunneling barriers, \cite{shakouri} therefore, the phonon conduction can be made negligible selecting barrier materials with low lattice thermal conductivity (in our model this could correspond to a strong reduction of damping rate $\gamma$). In any case, in this paper, we have pointed out that, even if one neglects the contribution from phonon thermal conductance, the electron-vibration coupling ($E_P$ in our model) is able to reduce the figure of merit. In order to improve the thermoelectric efficiency, molecules and metallic leads (which screen not only electron-electron but also electron-vibration interactions on the molecule) have to be selected to ensure a weak coupling between electronic and vibrational degrees of freedom.

Focus of the paper has been on the off-resonant electronic regime, where the thermoelectric properties show peak values, close to room temperature.
Obviously, the effects of Coulomb local interactions are expected to be negligible within these conditions. It would be interesting to extend the analysis to all the electron density regimes including the electron-electron interaction. Work in this direction is in progress. Finally, we point out that the electron-vibration interaction investigated in this paper is linked to the charge density injected by the external leads onto the molecule. Another possible source of coupling could come from the renormalization of the lead-molecule hopping integral induced by the center of mass movement. \cite{koch} Due to the large mass of the molecules investigated in this paper, we expect that the coupling through electron level density plays a major role.

%\section*{ACKNOWLEDGMENTS}

\begin{appendix}

\section{Generalized Langevin equation for the center of mass oscillator}

In this Appendix, we report the derivation of the Langevin equation for the molecule center of mass oscillator.

The resulting Langevin equation for the oscillator dynamics
\begin{equation}
m \frac{d v}{d t}=\xi(x,t)+F_{det}(x,v)
\label{langevin}
\end{equation}
has the position dependent fluctuating force term $\xi(x,t)$ and the deterministic force $F_{det}(x,v)$
\begin{equation}
F_{det}(x,v)=-k x + F^{el}(x,v)+ F_L^{ph}(x,v)+F_R^{ph}(x,v),
\label{fortot}
\end{equation}
where $F^{el}(x,v)$ is the force due to the effect of all the electronic degrees of freedom, and $F_{\alpha}^{ph}(x,v)$ is the force due to the coupling to the $\alpha$ lead phonon degrees of freedom.

Within the adiabatic regime, in Eq.(\ref{fortot}),  $F^{el}(x,v)=- \lambda N(x,v)$, where $N(x,v)$ is the electronic level occupation with an explicit dependence on the oscillator parameters. In order to make a self-consistent calculation, one needs the adiabatic expansion of the level occupation $N(x,v)$:
\begin{equation}
N(x,v) \simeq N^{(0)}(x)+v M^{(1)}(x),
\label{dens}
\end{equation}
where $N^{(0)}(x)$ is the zero order "static" term
\begin{equation}
N^{(0)}(x) = \frac{1}{2} \int_{-\infty}^{+\infty} \frac{ d (\hbar \omega)}{2 \pi}  [\sum_{\alpha} f_{\alpha}(\omega)] A(\omega,x),
\label{densz}
\end{equation}
with $A(\omega,x)$ given in Eq.(\ref{function}),  $f_{\alpha}(\omega)=1/(\exp{[\beta_{\alpha} (\hbar \omega-\mu_{\alpha})]}+1)$ free Fermi distribution of the $\alpha$ lead corresponding to the chemical potential $\mu_{\alpha}$ and the temperature $T_{\alpha}$ ($\beta_{\alpha}=1/k_B T_{\alpha}$), and $M^{(1)}(x)$
\begin{equation}
M^{(1)}(x) = -\hbar \lambda \int_{-\infty}^{+\infty} \frac{ d (\hbar \omega)}{2 \pi}
\frac{ \left[ \sum_{\alpha} f_{\alpha}(\omega) \right] (\hbar \Gamma)^2 I(\omega,x)}{\left[ I^2(\omega,x)+(\hbar \Gamma)^2/4 \right]^3},
\label{densu}
\end{equation}
with $I(\omega,x)=\hbar \omega - \epsilon -\lambda x$, is proportional to the first order "dynamic" contribution through the velocity $v$ and it is sensitive to charge fluctuations.
Therefore, using Eqs.(\ref{densz}) and (\ref{densu}), one gets $F^{el}(x,v)=-\lambda N^{(0)}(x) -A_{\lambda}(x) v $, with $A_{\lambda}(x)=\lambda M^{(1)}(x) $ positive definite position dependent dissipative term. In the regime investigated in this work, in Eq.(\ref{fortot}), one has $F_{\alpha}^{ph}(x,v)=-m \gamma_{\alpha} v$, with $\gamma_\alpha=\gamma/2$. Summarizing, the deterministic force $F_{det}(x,v)$
\begin{equation}
F_{det}(x,v)=F_{gen}(x)-A_{eff}(x) v,
\label{fortot1}
\end{equation}
consists of a generalized force $F_{gen}(x)$
\begin{equation}
F_{gen}(x)=-k x -\lambda N^{(0)}(x),
\label{fgen}
\end{equation}
and an effective position dependent damping term $A_{eff}(x)$
\begin{equation}
A_{eff}(x)=A_{\lambda}(x)+m \gamma.
\label{Aeff}
\end{equation}

In the adiabatic regime, exploiting the effect of the electron and phonon environment on the slow center of mass motion,
the fluctuating force $\xi(x,t)$ in Eq.(\ref{langevin}) is composed of three independent terms
\begin{equation}
\xi(x,t) = \xi^{el}(x,t) + \xi_L^{ph}(t) + \xi_R^{ph}(t),
\label{Langevin1}
\end{equation}
where $\xi^{el}(x,t)$ is due to the electronic degrees of freedom such that
\begin{equation}
\langle \xi^{el}(x,t) \rangle=0,\;\;\;\; \langle \xi^{el}(x,t) \xi^{el}(x,t') \rangle= D_{\lambda}(x) \delta(t-t'), \nonumber
\label{Langevin10}
\end{equation}
with \begin{equation}
D_{\lambda}(x) = \hbar \lambda^2 \frac{(\hbar \Gamma)^2}{4} \int_{-\infty}^{+\infty} \frac{ d (\hbar \omega)}{2 \pi}
\frac{ \sum_{\alpha,\eta} f_{\alpha}(\omega) \left[  1- f_{\eta}(\omega) \right] }{\left[ I^2(\omega,x)+(\hbar \Gamma)^2/4 \right]^2},
\label{densd}
\end{equation}
$\alpha,\eta=L,R$, and $\xi_{\alpha}^{ph}(t)$ is due to the $\alpha$ phonon lead such that
\begin{equation}
\langle \xi_{\alpha}^{ph}(t) \rangle=0,\;\;\;\; \langle \xi_{\alpha}^{ph}(t) \xi_{\alpha}^{ph}(t') \rangle= 2  K_B T_{\alpha} m \gamma_{\alpha}  \delta(t-t'). \nonumber
\label{Langevin1}
\end{equation}
Combining the three terms, one gets a fluctuating force $\xi(x,t)$ such that
\begin{equation}
\langle \xi(x,t) \rangle=0,\;\;\;\; \langle \xi(x,t) \xi(x,t') \rangle= D_{eff}(x) \delta(t-t'), \nonumber
\label{Langevin20}
\end{equation}
where the effective position dependent noise term $D_{eff}(x)$ is
\begin{equation}
D_{eff}(x)=D_{\lambda}(x)+ K_B (T_L+T_R) m \gamma,
\end{equation}

From the solution of the Langevin equation, the oscillator distribution function $Q(x,v)$ and the reduced position distribution function $P(x)$ are determined. In equilibrium conditions at temperature $T$, one has $P_{eq}(x)=C \exp{[-\beta V_{gen}(x)]}$, with C normalization constant, $\beta=1/(K_B T)$, and $V_{gen}(x)$ potential energy derived by the generalized force $F_{gen}(x)$ in Eq.(\ref{fgen}).

In order to calculate the thermal conductance, one can determine the vibrational energy currents directly from the derivative of the oscillator energy. \cite{pana} The oscillator is directly in contact with phonon leads, but only indirectly with  electron leads since there is the molecular level (see Fig.\ref{pisto}). In analogy with the terms in the deterministic (see Eq.(\ref{fortot})) and fluctuating (see Eq.(\ref{Langevin1})) force acting on the oscillator, the total energy current J involving the oscillator is composed of three terms \cite{brand}:
\begin{equation}
J =J^{el}_{\lambda}+J_L^{ph}+J_R^{ph},
\end{equation}
where $J^{el}_{\lambda}$ originates from the electron level and depends on the electron-vibration coupling
\begin{equation}
J^{el}_{\lambda}=\langle v \left[ \xi^{el}(x,t)- A_{\lambda}(x) v  \right] \rangle,
\end{equation}
and $J_{\alpha}^{ph}$ comes from the $\alpha$ phonon lead
\begin{equation}
J_{\alpha}^{ph}=\langle v \left[ \xi_{\alpha}^{ph}(t)- m \gamma_{\alpha} v  \right] \rangle.
\end{equation}
These quantities have to be evaluated along the dynamics. Once the stationary state is reached, the energy conservation requires that the total energy current J vanishes. Within the numerical simulations, we have not only found the total energy conservation, but that also $J^{el}_{\lambda}$ vanishes. Actually, the current mediated by the molecular electron level is not effective on the stationary state. Therefore, as emphasized in the main text, we have numerically calculated
the phonon thermal conductance in the linear regime. \cite{wang3,wang4}

\subsection{Weak electron-vibration coupling}
In the regime of weak electron-vibration coupling, the Langevin equation is simplified since the dissipative and fluctuating terms do not depend on the position x. Actually, in Eq.(\ref{fortot1}), $A_{eff}(x)$ is replaced by $A_{eff}$ since one gets $A_{\lambda}=\lambda M^{(1)}(x=0)$, with $M^{(1)}(x)$ given in Eq.(\ref{densu}). Therefore, within the weak coupling, there is a simple damping rate $\gamma_{\lambda}=A_{\lambda}/m$. Analogously, in Eq.(\ref{Langevin1}), $D_{eff}(x)$ is substituted by $D_{eff}=D_{\lambda}(x=0)$ in Eq.(\ref{densd}).

Within the weak-coupling regime, the Langevin equation is linear, therefore it can be analitically solved. The distribution functions $Q(x,v)$ and P(x,v) are Gaussian at and out of equilibrium. In particular, we have focused on the energy conservation finding analitically that not only $J=0$, but also $J^{el}_{\lambda}=0$ and $J_R^{ph}=-J_L^{ph}$. The quantity $G_K^{ph}$ has been explicitly evaluated as
\begin{equation}
G_K^{ph}=\frac{K_B \gamma (\gamma+\gamma_{\lambda})}{2} \int_{-\infty}^{+\infty} \frac{ d \omega}{2 \pi}
\frac{\omega^2}{(\omega_0^2-\omega^2)^2+\omega^2(\gamma+\gamma_{\lambda})^2}.
\label{gkfon}
\end{equation}

Therefore, there is a renormalization of $\gamma$ due to $\gamma_{\lambda}$, but not everywhere. We stress that, as expected, the direct link of $G_K^{ph}$ to $\gamma$ is still present. Moreover, as emphasized in the next Appendix, the order of magnitude of $\gamma_{\lambda}$ is always smaller than the values of $\gamma$ considered in this work within the weak coupling regime. Finally, we point out that the conductance $G_K^{ph}$ with $\gamma_{\lambda}=0$ represents a refencence value even when the electron-vibration coupling is strong, but the level occupation is very low. Actually, even in this case, the oscillator dynamics is not perturbed by the effects of electron-vibration interaction.

%\end{appendix}

%\begin{appendix}
\section{Electron-vibration damping rate and oscillator position distribution}

In this Appendix, we thoroughly discuss the features of the parameter $A_{\lambda}(x)/m$ in the linear response regime at the average temperature $T$. Moreover, in the same regime, we compare its $x$ dependence with that of the oscillator position distribution $P(x)$. The static distribution $P(x)$ is essentially the equilibrium distribution, so that it does not depend on the values of   $A_{\lambda}(x)/m$. However, a comparison of $x$ dependence between $A_{\lambda}(x)/m$ and $P(x)$ will clarify the conditions under which the electron-vibration interaction can affect the dynamics of the center of mass oscillator.
We will use the same parameters and units of the main text. Therefore $\hbar \Gamma \simeq 20$ meV will be the energy unit and $\Gamma$ the frequency unit. We will assume $\omega_0=0.25 \Gamma$ and the average chemical potential $\mu=0$. We will measure lengths in units of $\lambda / k$, times in units of $1/\Gamma$, temperatures in units of $\hbar \Gamma /K_B$. In this Appendix, we will consider the oscillator properties for $\gamma=0$ since we are interested on the effects induced by the electron-vibration coupling.

\begin{figure}[h]
\centering
\includegraphics[width=8cm,height=9.0cm]{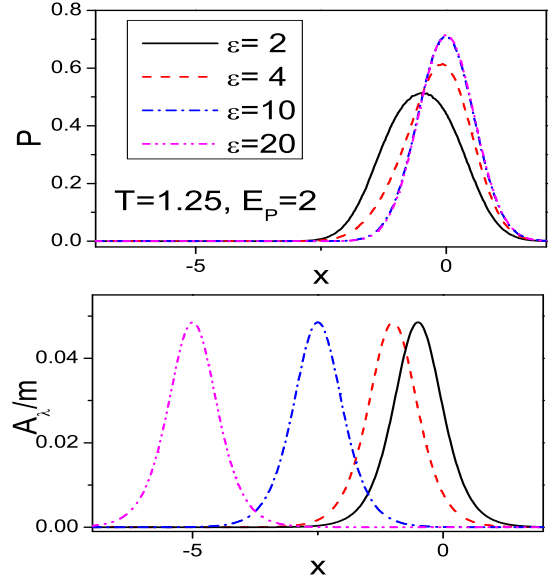}
\caption{(Color online) Upper panel: Oscillator position distribution $P(x)$ (in units of $k/ \lambda$) as a function of position x (in units of $\lambda / k$) for different values of level energy $\epsilon$ (in units of $\hbar \Gamma$) at temperature $T=1.25 \hbar \Gamma / k_B$ (close to room temperature) and
electron-vibration coupling $E_P=2 \hbar \Gamma$. Lower panel: Electron-vibration damping rate $A_{\lambda}(x)/m$ (in units of $\Gamma$) as a function of
position x (in units of $\lambda / k$) for different values of level energy $\epsilon$ (in units of $\hbar \Gamma$) at $T=1.25 \hbar \Gamma /k_B$ (close to
room temperature) and $E_P=2 \hbar \Gamma$.}
\label{figua1}
\end{figure}

As reported in the lower panel of Fig. \ref{figua1}, the peak values of $A_{\lambda}(x)/m$ are always smaller than the values of $\gamma$ considered in this paper ($ \gamma \simeq 0.15 -0.40 \hbar \Gamma$) even for the intermediate to strong value $E_P=2 \hbar \Gamma$ of the electron-vibration coupling. This means that the effects due to the electron-vibration coupling on the oscillator dynamics do not typically represent a large perturbation with respect to those induced by the coupling to phonon leads. Obviously, as reported in the figure, the effects of the electron-vibration coupling depends on the occupation of the electronic level. Indeed, the peak of $A_{\lambda}(x)/m$ largely shifts passing from the quasi-resonant case ($\epsilon=2 \hbar \Gamma$) to the off-resonant condition ($\epsilon=20 \hbar \Gamma$). In order to better quantify the effects of electron-vibration coupling on the oscillator dynamics, in the upper panel of Fig. \ref{figua1}, we report the oscillator position distribution $P(x)$ with varying the level energy $\epsilon$. In the quasi-resonant case ($\epsilon=2 \hbar \Gamma$), the peak positions of $A_{\lambda}(x)/m$ and $P(x)$ are almost coincindent. By the way, within the units considered in this paper, the peak position of the distribution $P(x)$ is about $- 2 n E_P/ \hbar \Gamma$, with $n$ density of the electron level. Therefore, with increasing $\epsilon$, the density $n$ is strongly reduced, so that the peak position of $P(x)$ quickly goes to zero. In the off-resonant cases ($\epsilon=10 \hbar \Gamma$ and $\epsilon=20 \hbar \Gamma$), in the linear regime considered in this Appendix, the distribution $P(x)$ is practically the Gaussian of the free harmonic oscillator at temperature T. On the other hand, in the off-resonant cases, the peak positions of $A_{\lambda}(x)/m$ go toward the direction opposite to the peaks of $P(x)$. In the regime of low occupation, the dynamics of the oscillator is not influenced by these effects for the value $E_P=2 \hbar \Gamma$, which is not negligible.

\begin{figure}[h]
\centering
\includegraphics[width=8cm,height=9.0cm]{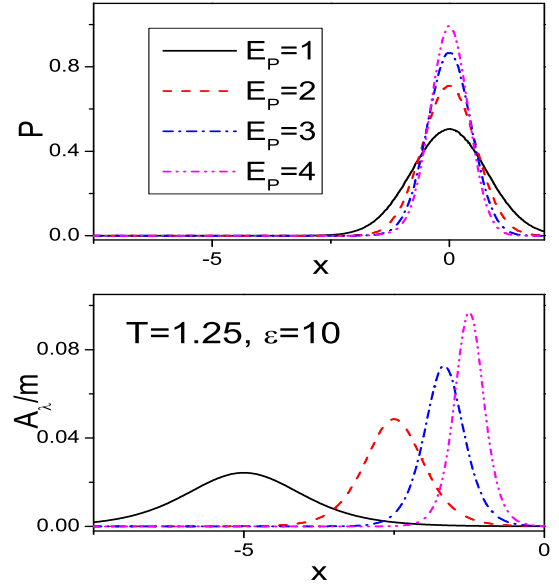}
\caption{(Color online) Upper panel: Oscillator position distribution $P(x)$ (in units of $k/ \lambda$) as a function of position x  (in units of $\lambda / k$)  for different values of electron-vibration coupling $E_P$ (in units of $\hbar \Gamma$) at temperature $T=1.25 \hbar \Gamma /k_B $ (close to room temperature) and level energy $\epsilon=10 \hbar \Gamma$. Lower panel: Electron-vibration damping rate $A_{\lambda}(x)/m$ (in units of $\Gamma$) as a function of position x (in units of $\lambda / k$) for different values of electron-vibration coupling $E_P$ (in units of $\hbar \Gamma$) at temperature
$T=1.25 \hbar \Gamma / k_B$) (close to room temperature) and level energy $\epsilon=10 \hbar \Gamma$.}
\label{figua2}
\end{figure}

In order to understand the interplay between the changes of $\epsilon$ and $E_P$, in Fig. \ref{figua2}, we have analyzed the behavior of $P(x)$ (upper panel) and $A_{\lambda}(x)/m$ (lower panel) for different values of $E_P$. We have considered the off-resonant case $\epsilon=10 \hbar \Gamma$. The distribution $P(x)$ is practically the Gaussian of the free harmonic oscillator at temperature T for different values of $E_P$ (it is so, since there is a change in the position unit with varying $E_P$). On the other hand, the peak positions of $A_{\lambda}(x)/m$ shift toward zero with increasing $E_P$. Only for the large coupling $E_P=4 \hbar \Gamma$, the superposition between $P(x)$ and $A_{\lambda}(x)/m$  is not negligible.
\end{appendix}

\begin{acknowledgments}
The authors would like to thank financial support from the project GREEN (PON02-00029-2791179) funded by Ministero dell'Istruzione, dell'Universit\`a e della Ricerca.
\end{acknowledgments}

%\addcontentsline {toc}{chapter}{Bibliografia}

\end{document}